\documentclass[12pt,preprint]{aastex}

\shorttitle{Star-Disk Connection in Rotational Evolution}
\shortauthors{Kundurthy et al.}

\begin{document}
\title{Mid-IR Observations of T Tauri stars: Probing the Star-Disk Connection in Rotational Evolution}

\author{Praveen Kundurthy and Michael R. Meyer}
\affil{Steward Observatory, The University of Arizona, Tucson, AZ 85721}
\email{pkundurthy@as.arizona.edu, mmeyer@as.arizona.edu}

\author{Massimo Robberto and Steven V.W. Beckwith}
\affil{Space Telescope Science Institute, Baltimore, MD 21218}
\email{robberto@stsci.edu, svwb@stsci.edu}

\author{Tom Herbst} \affil{Max-Planck-Institut f\"{u}r Astronomie,
Heidelberg D-69117, Germany} \email{herbst@mpia-hd.mpg.de}

\begin{abstract}
We present mid-IR N-band $(\lambda_{eff} = 10.2\micron)$ photometry of a carefully selected sample of T Tauri stars thought to be single from the Taurus-Auriga molecular cloud. Infrared excesses in these stars are generally attributed to circumstellar dust-disks. Combining observations at 2.16$\micron$ (K$_{s}$-band) and 10.2$\micron$ (N-band) we probe a region in the circumstellar dust-disk from a few stellar radii through the terrestrial planet zone (0.02-1.0AU). By analyzing the distribution of the $(K_{s}-N)$ color index with respect to previously measured photometric rotation periods we investigate what role circumstellar disks play in the rotational evolution of the central star. The resulting positive correlation between these two variables is consistent with the notion that a star-disk interaction facilitates the regulation of angular momentum during the T Tauri stage. We also demonstrate, how including non-single stars in such an analysis will \textit{weaken} any correlation in the relation between $(K_{s}-N)$ color and period. To further understand disk properties we also present SEDs for a few objects with new ground based M-band $(\lambda_{eff} = 4.8\micron)$ and Q-band $(\lambda_{eff} = 20\micron)$ data and compare them to a geometrically thin, optically-thick disk model.
\end{abstract}


\section{Introduction}
As a rotating molecular cloud core collapses to form a star it must shed vast amounts of angular momentum to explain the observed differences between the specific angular momentum of molecular cloud cores and stars on the main-sequence \citep{bod95}. Various physical processes, which serve as solutions to this ``angular momentum problem'' are thought to occur as an object transitions from a cloud core to a Zero-Age Main Sequence (ZAMS) star. For low-mass stars in the pre-main sequence stage (T Tauri stars) the regulation of stellar rotational angular momentum is thought to be facilitated by the formation of and subsequent interaction with a circumstellar disk. Circumstellar disks are thought to be common by-products of the star formation process \citep{BS96}. They are easily identified by peculiar features like infrared and ultraviolet excesses, and in some rare circumstances can also be detected through direct imaging \citep{Odell93,MO96}. \citet{KH95} estimate $> 50\% $ of stars in the \object{Taurus-Auriga} star forming region have circumstellar disks, while \citet{hillenbrand98} estimate that $> 80\%$ of stars in the Trapezium cluster possess circumstellar accretion disks \citep[see also][]{haisch01}. The IR excesses from these disks are the result of heating circumstellar gas and dust by radiation from the central star and through viscous accretion. The disk temperature distribution is often modeled as a power-law. If the disk is optically thick it will radiate as a blackbody with the shortest wavelengths dominated by higher temperature material in the inner disk and the longer wavelengths dominated by cooler material in the outer regions \citep{beckwith99}. Thus, emission at different wavelengths can be used to probe different regions in the disk.

The pioneering study by \citet{edwards93} for a set of stars from the Taurus, Auriga, Chameleon, Orion and Lupus star--forming regions, compared stellar rotation periods with the $(H-K)$ color index. This color index probes a region in the disk that lies within a few stellar radii $(\sim 10R_{\sun})$. They found stars whose $(H-K)$ colors indicated the presence of an accretion disk were rotating more slowly than stars whose $(H-K)$ colors indicated the absence of an accretion disk. A similar study by \citet{bouvier93} suggested a bimodal distribution of slowly rotating classical T Tauri stars (cTTS, showing evidence of active accretion from a circumstellar disk) and fast rotating weak T Tauri stars (wTTS, weak-lined $H\alpha$ emission-line stars with no signs of accretion). Studies of the \object{Orion Nebula Cluster} (ONC) have yielded mixed results. \citet{CH96} report a bimodal distribution in photometric periods of T Tauri stars in Orion as evidence for the disk-locking phenomena \citep[see also][]{herbst02}. Theoretical models have always relied heavily on some form of direct or indirect angular momentum transfer between stars and their disks to explain the redistribution of rotational angular momentum during the transition from the Pre-Main Sequence (PMS) stage to the ZAMS stage. However, an extensive study conducted by \citet{stassun99,stassun01} report no bi-modality in period distribution and no dichotomy of disked slow rotators and disk-less fast rotators. A recent review of this literature can be found in \citet{mathieu04}. In this study we use the $(K_{s}-N)$ color index to probe a region in T Tauri dust-disks that extends from a few stellar radii through the terrestrial planet zone (0.02-1.0AU). Comparing the $(K_{s}-N)$ color index of carefully selected single stars with previously measured rotation periods allows us to investigate any effects the disk might have on the angular momentum of the star.

This paper outlines an observational study that expands on the work presented in \citet{MB97}, and hopes to further our understanding of the role of disks in the rotational evolution of T Tauri stars. In section 2 we present a description of our sample, the observations and the data reduction procedure. In section 3 we present the relationship between the $(K_{s}-N)$ color index and rotation period of single PMS stars in Taurus-Auriga, and the spectral energy distributions (SEDs) of a few stars with new M and Q-band data. In section 4 we discuss the results in light of current theories for the rotational evolution of T Tauri star and disk systems. We also demonstrate that including binaries in a study of $(K_{s}-N)$ vs. rotation period diminishes any observed correlation. We conclude with a summary in section 5.
 
\section{Observations and Data}
We selected stars in the \object{Taurus-Auriga} SFR, that have been surveyed for, but are known to be lacking companions from the multiplicity surveys of \citet{ghez93}, \citet{leinert93} and \citet{simon95}, that also have photometric period data available in the literature. Photometric period data for all stars were taken from \citet{bouvier93, bouvier95} and \citet{osterloh96}. Infrared K$_{s}$-band data for all stars were taken from the Two-micron all sky survey (2MASS), point-source catalog. The typical magnitude uncertainties for the K$_{s}$-band data ranged between $2\%-3\%$ for the stars of our sample. We obtained new N-band observations for 12 stars. The details of the observations, such as filter information, exposure times, and flux standards are listed in Table 1. We selected 18 more stars that satisfied the selection criterion above and also had N-band measurements in the literature with photometric uncertainties $\leq 30\%$. The N-band magnitudes were queried in the catalog \citet{gezari99} and references therein. The original literature source of N-band magnitudes are noted in the last column of Table 2. When multiple measurements of an object were returned from \citet{gezari99} we chose the most recent measurement or the measurement with the smallest magnitude uncertainty. Table 2 also lists spectral types, T Tauri types, SED classes, photometric periods, effective temperatures and luminosities for all 30 objects in our sample \citep{bouvier93,bouvier95,osterloh96,strom89,KH95}. The sample covers an age range of 0.1 - 10Myrs, and a mass range of $0.2 - 2.0 M_{\sun}$. The ages and masses were estimated from the tracks of \cite{DM97} and typically have uncertainties of order $\times$ 2. The ratio of CTTS to WTTS in our sample is 14:15. The vast majority of objects in our sample are optically visible Pre-Main Sequence stars with class II or class III SEDs and well-determined spectral types. However, we do include a heavily veiled object FT Tau, a continuum star of SED class II.

We also made N-band observations of the binary star UX Tau (see Table 1). M and Q-band observations (see Table 1) for a few single stars were also made in order to derive SEDs that might illuminate properties of circumstellar disks surrounding these individual sources.

\subsection{Observations}
M, N and Q-band photometric observations were made using the Mid-infrared Array eXpandable (MAX; \citet{RH98}) camera at the 3.8m United Kingdom Infrared Telescope (UKIRT) on the nights of February 6th and 8th, 1997. The MAX camera was built by Infrared Laboratories of Tucson and contains a 128$\times$128 pixel Rockwell HF16 Si:As BIB detector. The plate-scale of the detector was 0.265$\arcsec$/pixel and corresponded to a field size of $33.9\arcsec\times33.9\arcsec$ on UKIRT. The typical observing mode of chopping and beam switching were utilized to take images of standard stars and program stars with matched chopping and beam-switching amplitudes of approximately 10.3$\arcsec$. Spatial and temporal variations of the sky emission in the mid-IR were reduced by this observational procedure. Typical integration times ranged from 10.24-20.48 seconds for all bands. These chop-subtracted frames were co-added and the resulting total integration times are listed in Table 1. The typical read noise and gain values for this detector were 1250 e- and 400 e-/ADU respectively.

\subsection{Data Reduction}
Each chop-subtracted frame contained two signals, one negative and one positive, of the source placed approximately 10.3$\arcsec$ apart. The frames were cleaned for cosmic rays using a 3-sigma rejection procedure and then combined through subtraction of beam-switched pairs of similar frames of the same object to create a final image with one central positive signal and two satellite negative signals. Additional bad pixels were identified by eye and interpolated over using adjacent pixels with FIXPIX in IRAF\footnote{IRAF is distributed by the National Optical Astronomy Observatories, which are operated by the Association of Universities for Research in Astronomy, Inc., under cooperative agreement with the National Science Foundation.}. The PHOT task in IRAF's NOAO.DIGIPHOT.APPHOT package was used to derive aperture photometry on the central signals in each object frame. The optimal aperture was selected such that the combined uncertainties due to PSF fluctuations (resulting in uncertainties in aperture corrections) and uncertainties in the sky from larger apertures were minimized. The standard stars listed in the last column of Table 1 were used to determine the optimal apertures for each band. For the N-band, a target aperture of R = 2.65$\arcsec$ (10 pixels) was used. A sky-annulus to determine the mean residual background was chosen such that significant source flux was excluded and the error in mean sky (within the annulus) was minimized. For the N-band the sky annulus used extended from 5.035$\arcsec$-5.698$\arcsec$ (19 - 21.5 pixels). Photometric precision ranged from $3\%-9\%$ for most targets in the N-band except for \object{HD 283572} and \object{LkCa 21} with 18$\%$ and 25$\%$ photometry respectively. Magnitude upperlimits were derived for non-detections \object{TAP 26}, \object{TAP 40}, \object{TAP 41} and \object{V830 Tau}. We assumed a minimum object area of 4 pixels$(2\times2)$ to estimate a $3\sigma$ flux upper-limit in each respective non-detection frame. These upper limits correspond to unusually large source fluxes due to the high sky-noise in the non-detection frames. Similar photometric techniques were used for the M-band and Q-band data. The M-band target aperture was R = 2.915$\arcsec$ (11 pixels) with a sky annulus from $5.035\arcsec-5.963\arcsec$ (19 - 22.5 pixels). The Q-band target aperture was R = 1.855$\arcsec$ (7 pixels) with a sky annulus from $2.915\arcsec-7.155\arcsec$ (11 - 27 pixels). For comparison, the diffraction limited diameters of the 3.8m UKIRT beam at wavelengths of interest are 0.33$\arcsec$, 0.66$\arcsec$ and 1.3$\arcsec$ at 4.8$\micron$, 10.2$\micron$ and 20$\micron$ respectively. Zero-point fluxes for the standard stars in the filters used were taken from \citet{cohen92}.  The N--band photometry is presented in Tables 2 and 4 for 12 single T Tauri stars and one binary, UX Tau A.  M--band photometry for two stars, and Q--band photometry for five stars is presented in Table 3. 

\section{Results}
To investigate whether there is a trend of disked slow rotators and disk-less fast rotators among T Tauri stars, we plot the $(K_{s}-N)$ color index against the photometric rotation period of single T Tauri stars from the Tau-Aur molecular cloud in Figure 1a. We tested for a correlation between $(K_{s}-N)$ and period using a linear least-squares analysis on the data of Figure 1a. A first order polynomial was fit to the data. The fit resulted in a correlation coefficient of 0.52 for 25 data points (excluding 5 non-detections). This indicates that the probability the variables in this linear fit are uncorrelated is $< 0.009$. In order to remove scatter in the correlation, we repeated this analysis explicitly taking into account the intrinsic colors and extinction of each source. We adopted the infrared extinction law from \citet{cohen82} and used visual reddening values (A$_{v}$) reported by \citet{KH95} for our sample of objects. We estimated intrinsic $(K_{s}-N)_{o}$ values from \citet{mamajek04}. Figure 1b shows the quantity $E(K_{s}-N)_{o}$ plotted against the rotation period. Where $E(K_{s}-N)_{o} = (K_{s}-N) - 0.06A_{v} - (K_{s}-N)_{o}$, is the difference between reddening corrected color observed and intrinsic color (cf. \citet{meyer97a}). A linear least-squares analysis results in a correlation coefficient of 0.48 for 25 data points. The probability that these variables are uncorrelated is $<0.016$. This is consistent with the previously derived correlation.  A positive correlation could be interpreted as evidence for disk-assisted regulation of stellar angular momentum among single T Tauri stars in Taurus-Auriga. It is important to note that the observed magnitude uncertainties were not used to weight the fit since they represent photometric uncertainties from many non-simultaneous observations in the K$_{s}$ and N-bands. These are much smaller than the expected range of variability in the infrared \citep{rydgren84,skrutskie96} which presumably contributes to the observed scatter in the correlations. Figure 1a and 1b are marked by horizontal dashed lines that designate important limits in the distributions of the $(K_{s}-N)$ and $E(K_{s}-N)_{o}$ color indices. The line at $(K_{s}-N) = 2.0m$ in Figure 1a represents the expected lower limit for an excess from a face-on, optically--thick and geometrically--thin disk \citep{hillenbrand92}. Taking into account the extinction, for the objects in our sample, the same limit falls at $E(K_{s}-N)_{o} = 1.8m$ in Figure 1b. The dashed line in Figure 1a at $(K_{s}-N) = 1.0m$ marks the expected maximum color for a normal reddened photosphere, and the line in Figure 1b at $E(K_{s}-N)_{o} = 0.84m$ marks the value of a 3-sigma excess from the dispersion of intrinsic colors for stars in our sample \citep{mamajek04}. This is the lower limit for the detection of a disk. 

Figure 2 shows the histogram of $(K_{s}-N)$ values of the stars in our sample separated into bins of 0.5\textit{m}. There is a paucity of stars between 1.0m $<$ $(K_{s}-N)$ $<$ 2.0m. This confirms the findings of \citet{S90} \citep[see also][]{SP95,KH95,WW96} that so-called ``transition objects'' within this color range are rare. It suggests that the time to evolve from optically-thick to optically-thin in the 0.02-1.0 AU region is short compared to the average age of T Tauri stars (i.e. $\ll$ 1 Myr). The only star within this region is DH Tau ($(K_{s}-N)$ = 1.92m and $E(K_{s}-N)_{o}$ = 1.53).  

We classified objects from Figures 1a and 1b into disked and disk-less stars and compared their period distributions. Objects with $(K_{s}-N) < 1.0m$ or $E(K_{s}-N) < 0.84m$, were classified as disk-less and objects with $(K_{s}-N)$ and $E(K_{s}-N)_{o}$ greater than these respective limits were classified as disked. The two sets contained the same objects for Figure 1a and 1b. We found stars with disks rotate slower on average $(P_{avg} = 7.46d)$ and stars without disks rotate faster $(P_{avg} = 4.30d)$. The period distributions are wide for both ($\sigma$ = 2.44 and 2.57 for disked and disk-less respectively). A two sided K-S test of the period distribution of disked and disk-less stars resulted in D = 0.68. Which indicates the probability that the period distributions are drawn from the same parent is $1.6\%$. In the language of gaussian hypothesis testing, this is a 2.4$\sigma$ result. So for stars which are believed to be single, the period distributions of disked and disk-less stars, are marginally different.

As described in section 2, we obtained M-band and Q-band photometry for a few stars with strong IR excesses. All new M and Q-band photometry are listed with other photometry assembled from the literature in Table 3. SEDs derived from these data are shown in Figure 3. The data are not simultaneous. We compared the de-reddened fluxes to emperical stellar photospheres appropriate for each spectral type and models of geometrically-thin, optically-thick disks \citep{hillenbrand92}. The model is of a face-on blackbody disk extending from the stellar surface. Easily visible in all SED plots are signs of accretion as the excess fluxes in the U-band $(\lambda_{eff} = 0.36\micron)$. IR excesses are also present and significantly exceed the predictions of the stellar photosphere and the flat reprocessing disk model at longer wavelengths. Most sources lack IR excesses at the shortest wavelengths, consistent with dust sublimation and/or inner holes (\citet{muzerolle03}; cf. \citet{meyer97a}). An inner dust disk hole would be visible as a lack of near-infrared excess in the SED. Disk regulation theories predict that circumstellar disks have inner holes maintained by pressure balance with the stellar magnetic field \citep{OS95}. Measurements of stellar magnetic field strengths are needed along with estimates of $M_{\star},\dot{M}$ and R$_{\star}$ to properly estimate the extent of inner disk clearing. The very large IR excesses at longer wavelengths indicate that the disks are probably not flat but flared \citep{CG97} with vertical temperature structures commonly referred to as ``disk atmospheres'' \citep[see][]{dalessio99,dullemond01}.

\section{Discussion}
In the following sections we discuss the significance of our results in the context of current theoretical frameworks for rotational evolution. We also compare our observational study to others conducted in Tau-Aur and the ONC. Finally we emphasize the importance of sample selection criteria for studying rotational evolution of low mass PMS stars.

\subsection{Disk-regulation}
The positive correlation, at the 99$\%$ confidence level, between the $(K_{s}-N)$ color index and rotation period supports the idea that disks play a role in regulating the angular momentum of T Tauri stars. Most models of PMS angular momentum evolution rely on some form of disk-assisted angular momentum regulation. In the absence of a disk, we would expect young stars to conserve total angular momentum as they contract towards the ZAMS, spinning-up as their radii decrease over time. Our results suggest that the disk is countering the spin-up effect caused by ordinary stellar contraction and to some extent the spin-up torque caused by the accretion of high angular momentum material from the disk boundary \citep{GL79a,GL79b}. Evidence for accretion is seen in the U-band excess (Figure 3) of even the fast rotator \object{GK Tau}. \citet{camenzind90}, \citet{konigl91} and \citet{CC93} explore the idea of magnetospheric star-disk coupling. In this model the stellar magnetic field is thought to couple with the disk, extending to regions where the disk is spinning slower than the stellar surface. As a result the star is braked and eventually slows its rotation significantly. \citet{shu94} and \citet{OS95} suggest it is unlikely for magnetic coupling to occur over extended regions of the disk. Their models allow for the disk to be penetrated by the stellar magnetic field only near the co-rotation radius, hence making the transfer of angular momentum via rotational coupling insignificant. Instead, angular momentum is lost from the stellar surface through the expulsion of X-winds (eXtraordinary winds) near the star-disk boundary. The magnetized X-winds are driven by accretion from the disk. Similarly \citet{vRB04} and \citet{MP05} explore a scenario where angular momentum is lost via accretion-powered stellar winds. In all theoretical frameworks (disk-locking, X-winds or stellar winds) the presence of an extended disk (0.02-1.0AU) is vital to angular momentum regulation. The magnetospheric disk locking models require an extended magnetized disk beyond the co-rotation radius to provide a spin-down torque. The wind models require the presence of accretion from the inner disk, which in turn depends on the extended disk to persist. 

T Tauri stars are thought to dissipate their disks over a range of times during their evolution \citep{strom89,S90,haisch01}. We conducted a two-sided Kolmogorov-Smirnov(K-S) test to see if the age distributions for disked and disk-less stars (as defined by the $(K_{s}-N)$ index) in our sample are similar. We derived a D-statistic of 0.32 indicating that the age distributions are consistent with having been drawn from the same parent population. If disk-assisted regulation is a dominant force the angular momentum evolution of a T Tauri star, then the fact that stars dissipate their disks at different times would explain the wide dispersion of rotational rates observed for stars on the ZAMS \citep{BFA97,K97}. In other words stars that dissipate their disks early are believed to spin-up and become rapid rotators, while stars that dissipate their disks late, end up as slow rotators \citep{bouvier94,rebull04,HM05}.

\subsection{Sample Selection}
Most rotational evolution models explore only the case of a single star's interaction with its disk. Comparing the $(K_{s}-N)$ index with the rotation period of single stars enables us to compare our results to the theoretical frameworks which use single star + disk scenarios. Furthermore, there is some evidence that the presence of a companion affects the evolution of a disk around a PMS object \citep{meyer97b,ghez94}. \citet{jensen96} and \citet{JM97} present convincing evidence that disks from 50-100AU are affected by the presence of binary companions over this range of separations. 

In Figure 4 we combine data points from Figure 1 with $(K_{s}-N)$ and period data available for unresolved binaries (listed in Table 4). Performing a linear correlation test similar to the one described in section 3, for $(K_{s}-N)$ vs. rotation period, resulted in a correlation coefficient of 0.35, for 38 data points (not including upper-limits). This means the probability that linear correlation is random is $< 0.05$, suggesting that the presence of unresolved binaries in a sample of young stars could \emph{weaken} any expected correlation. We also conducted a two sided K-S test on the objects in Figure 4, separating them into disked and disk-less stars, as described in section 3. We derived a D-statistic of 0.45, which indicates that the probability the distributions are similar is $7.0\%$. This is a $1.0\sigma$ result. So we cannot confidently assert that the period distributions were drawn from different parent populations. \citet{ghez93} estimate that 2/3 of the stars in the Tau-Aur region are multiple systems. Yet, the linear correlation test only includes 13 binary systems compared to the 25 single systems (excluding upperlimits). This is because there is a lack of photometric period data for approx 85$\%$ of the stars surveyed to be multiple systems, compared to 50$\%$ for those survyed to be single systems \citep{ghez93,leinert93,simon95,bouvier93,bouvier95,osterloh96}. This is probably due to the difficulty in interpreting data from binary systems. Most photometric period surveys that include binaries do not resolve companions and hence it is hard to know for certain a) which component's period was determined (if the stars are of comparable brightness) or b) if the light fluctuations were due to oribital motions (eclipsing) or actual rotational motion (spots). In addition, components in a close binary system are hard to resolve in the infrared. The source of N-band flux from an unresolved multiple system cannot be easily determined. Contributions from each stellar component, the dust disk around each object, or a circumbinary disk \citep{ghez94} could add to the N-band flux. In investigating star-disk interactions in the PMS, it is important to restrict the sample to well-studied single stars, or resolved binaries.

\subsection{Resolving Apparent Discrepancies}
The results from \citet{edwards93} for the $(H-K)$ color index shows evidence for disk-less fast rotators and disked slow rotators in Tau-Aur.  However, the $(H-K)$ color index probes a region in the dust-disk very close to the stellar surface (approx $<0.02$ AU). \citet{muzerolle03} estimate that the typical temperature at which dust sublimates around T Tauri stars is approximately 1400K, which roughly corresponds to the temperatures of the hottest dust emitting in K$_{s} (\lambda_{eff} = 2.16\micron)$ and L-band $(\lambda_{eff} = 3.54\micron)$. So emission from the dust-disk is limited to the region with $T_{disk} < 1400K$. If we consider a case where the disk truncation radius lies far beyond the sublimation zone (where $T_{inner-disk} << 1400K)$ at $r_{inner} >> 0.02 AU$, the disk might still be effectively disk-locked but will not display any excess in H and K-bands. There are 8 stars from our single star set which are also studied by \citet{edwards93}. Interestingly, 1 of the 8 stars (\object{DN Tau}) lacks a $(H-K)$ excess but \textit{does display a} $(K_{s}-N)$ \textit{excess} (P$_{rot}$ = 6.0 days). This could mean that its disk is truncated outside of the sublimation radius. The other 7 stars either lack both $(H-K)$ and $(K_{s}-N)$ excesses or display both $(H-K)$ and $(K_{s}-N)$ excesses. There are \textit{no cases} where an object lacks a $(K_{s}-N)$ excess but shows an $(H-K)$ excess.

In a complementary study, \citet{stassun99} surveyed a large sample of stars in the ONC at near-IR wavelengths, with periods $< 8.0$ days, and discovered no bimodality in the distribution of rotation periods and no correlation of (I-K) color excess with rotation periods. However, subdividing the sample by mass to select only stars $> 0.25M_{\sun}$ demonstrated that the higher-mass PMS stars display the bimodal distribution of rotation periods \citep{herbst00}. The \citet{stassun99} sample is dominated by very low-mass stars which may possess different magnetic star-disk coupling mechanisms. \citet{CB00} argue that the ONC population differs significantly from the \object{Taurus} population in that PMS stars in Taurus are generally slow rotators. \citet{hartmann02} suggests that this could be because the typical disk-braking time-scale of T Tauri stars is comparable to the ages of T Tauri stars in the ONC. Hence, the mechanism might not have had time to brake most disked stars in the ONC. \citet{stassun01} obtained mid-IR data for a sample of stars from the ONC and Tau-Aur selected not to have near-IR excesses. These data indicate no trend of slow rotators with mid-IR excesses in their sample for both clusters. \citet{stassun01}, like most previous studies, include single, binary and stars that have not been surveyed for companions in the analysis that led to a null result. Interpretation of period and infrared fluxes of binary stars is not straightforward as discussed in the previous subsection. In fact, for the single stars in their sample (also included in our sample) their results are entirely consistent with ours.

Some of our results echo the observations made by \citet{stassun01} that there is no strict dichotomy of disked slow rotators and disk-less fast rotators. A handful of stars qualify as abnormal cases. \object{GK Tau} is a fast rotator with a period of 4.65 days. Yet it displays a strong infrared excess, $E(K_{s}-N) = 3.20$. The well-known PMS star \object{SU Aur} appears to be a disked $(E(K_{s}-N) = 3.33)$ fast rotator (P = 3.50d) and \object{IP Tau} is definitely a disked $((K_{s}-N) = 2.30)$ fast rotator (P = 3.25d). \object{LkCa 21} is a disk-less slow rotator and \object{V819 Tau} is disk-less ``moderate-to-fast'' rotator, classified as a ``transition object'' by \citet{S90}. We queried 2MASS H and K$_{s}$-band magnitudes and derived the $E(H-K_{s})_{o}$ index for these 5 objects. $4/5$ objects have near-IR properties similar to their mid-IR $E(K_{s}-N)_{o}$ properties. GK Tau, SU Aur and IP Tau have significant $E(H-K_{s})_{o}$ excesses while LkCa 21 has no $E(H-K_{s})_{o}$ excesses. Perhaps LkCa 21 has very recently lost its disk and has not yet had time to spin up (cf. DI Tau; \citet{meyer97b}).  Of these ``special case'' objects, only V819 Tau is included in the study by \citet{edwards93}. It has no near--IR excess, a marginal (K-N) excess, and a period of 5.6 days.  Including all these objects in our analysis, we see a dispersion in the $E(K_{s}-N)_{o}$ vs. rotation period plot comparable to that seen in the $E(H-K)_o$ analysis of \citet{edwards93}. It would be interesting to further investigate the physical properties (M$_{\star}$,$\dot{M}$, R$_{\star}$,B) of these objects that are expected to determine the co-rotation point in their disks. For example, if $\dot{M}$ is large enough in the protostellar phase, the inner disk might crush the magnetosphere resulting in a population of rapidly rotating objects \citep{najita95,covey06}. Note that the disked fast rotators here have moderate to low accretion rates \citep{GHBC98}. A sample of stars with a fixed M$_{\star}$,R$_{\star}$, and B should exhibit a correlation between P$_{rot}$ and $\dot{M}$ in the x-wind model. Such studies might also elucidate the \emph{dispersion} in the $(K_{s}-N)$ vs. period graph. 

From a theoretical perspective, we don't necessarily expect a linear correlation between $(K_{s}-N)$ and P$_{rot}$. However, our fit demonstrates the low probability that these variables are uncorrelated. Our two-sided K-S test of the periods of disked vs. disk-less stars also hints at a correlation. \citet{rebull04} point out that correlations between P$_{rot}$ and dust-disk or accretion indicators might be difficult to establish given that the spin-up time is of order the age of typical pre-main sequence samples \citep{hartmann02}. The fact that we do see a correlation and that our disked and non-disked stars appear to be of similar age (1-3 Myr) suggests that these weak-emission T Tauri stars in Taurus lost their disks very early in their pre-main sequence evolution. Deep far-infrared or millimeter observations might yet reveal whether they retain any disk signatures at all (e.g. \citet{AW05}). New surveys for disks at all radii enabled with the Spitzer Space Telescope combined with photometric rotation studies will shed considerable light on these remaining questions.

\section{SUMMARY}
$1.$ Analysis of the $(K_{s}-N)$ vs. rotation period data for single T Tauri stars in Tau-Aur cloud suggests a correlation implying that disked stars are generally rotating slower than disk-less stars. We selected stars that have been surveyed to be single and have documented photometric rotation periods and available N-band fluxes. There are a few abnormal cases which cause a dispersion in the $(K_{s}-N)$ vs. period relation and would be interesting candidates for further study.

$2.$ A histogram of $(K_{s}-N)$ colors confirms the result from \citet{S90}, \citet{SP95}, \citet{KH95} and \citet{WW96} that stars between $1.0m < (K_{s}-N) < 2.0m$ are rare. The transition from optically thick to optically thin must be much less than the average lifetime of T Tauri stars ($\ll$ 1 Myr).

$3.$ SEDs of a few classical T Tauri stars, including new mid-infrared photometry, confirm that most of these stars are accreting actively from their circumstellar disks. IR excesses are significantly greater than fluxes predicted by the flat-reprocessing disk model of \citep{hillenbrand92}. This supports the idea that disks are not flat but are flared and probably have atmospheres \citep{CG97,dalessio99,dullemond01}.

$4.$ Analyzing Mid-IR fluxes and period data for single stars are best for investigating star-disk rotational interaction because: 1) It is hard to resolve the components of a binary system in the Mid-IR, and 2) photometric period data for binaries are ambiguous. Including binaries can statistically \emph{weaken} any observable correlation caused by star-disk interactions in a population of single stars.

\acknowledgements
We would like to thank Peter Biezenberger, Christoph Birk, and the staff of UKIRT for their help in commissioning the MAX camera. We would like to thank S. Edwards, S. Wolff and an anonymous referee for helpful comments that improved this manuscript. This material is based on work supported by NASA through the NASA Astrobiology Institute under cooperative agreement CAN-02-OSS-02 issued through the Office of Space Science. This paper makes use of data products from the Two Micron All Sky Survey, which is a joint project of the University of Massachusetts and the Infrared Processing and Analysis Center/California Institute of Technology, funded by the National Aeronautics and Space Administration and the National Science Foundation.

{\it Facilities:} \facility{United Kingdom Infrared Telescope(UKIRT)}.

\clearpage

\begin{figure}
\plotone{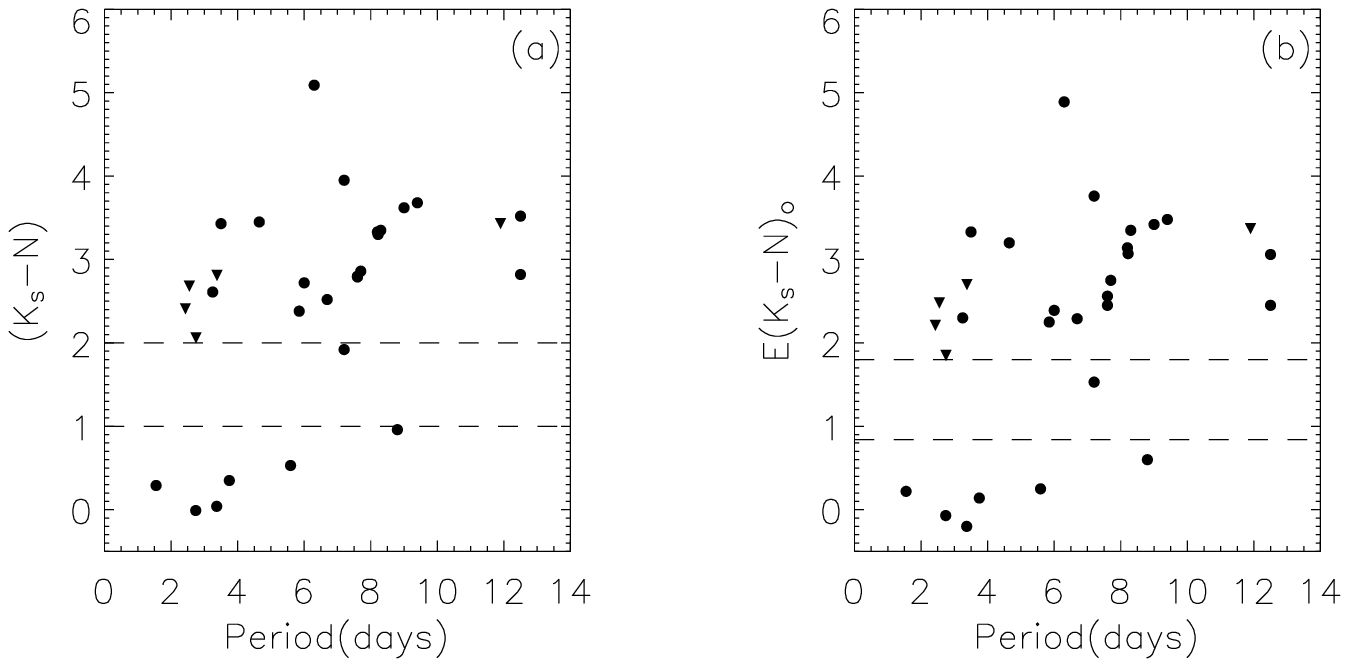}
\caption{Figures 1a and 1b are graphs of $(K_{s}-N)$ vs. period and $E(K_{s}-N)_{o}$ vs. period, of a sample of stars from Tau-Aur that are thought to be single. The lower limit for an optically thick circumstellar disk $(K_{s}-N) = 2.0m$ \citep{hillenbrand92} and the upper limit for photospheric emission plus extinction $(K_{s}-N) = 1.0m$ are indicated by the dashed lines on Figure 1a. The extinction corrected lower limit for an optically thick disk is marked by the dashed line at $E(K_{s}-N)_{o} = 1.8m$ on Figure 1b. Also on Figure 1b, the dashed line at $E(K_{s}-N) = 0.84m$ marks the typical value of a 3-$\sigma$ excess (lower limit for the detection of a disk) for the stars in our sample \citep{mamajek04}. The downward pointing triangles represent 
upper limits in $(K_{s}-N)$ and $E(K_{s}-N)_{o}$ above the dispersion in intrinsic color.\label{fig1}}
\end{figure}
\clearpage

\begin{figure}
\plotone{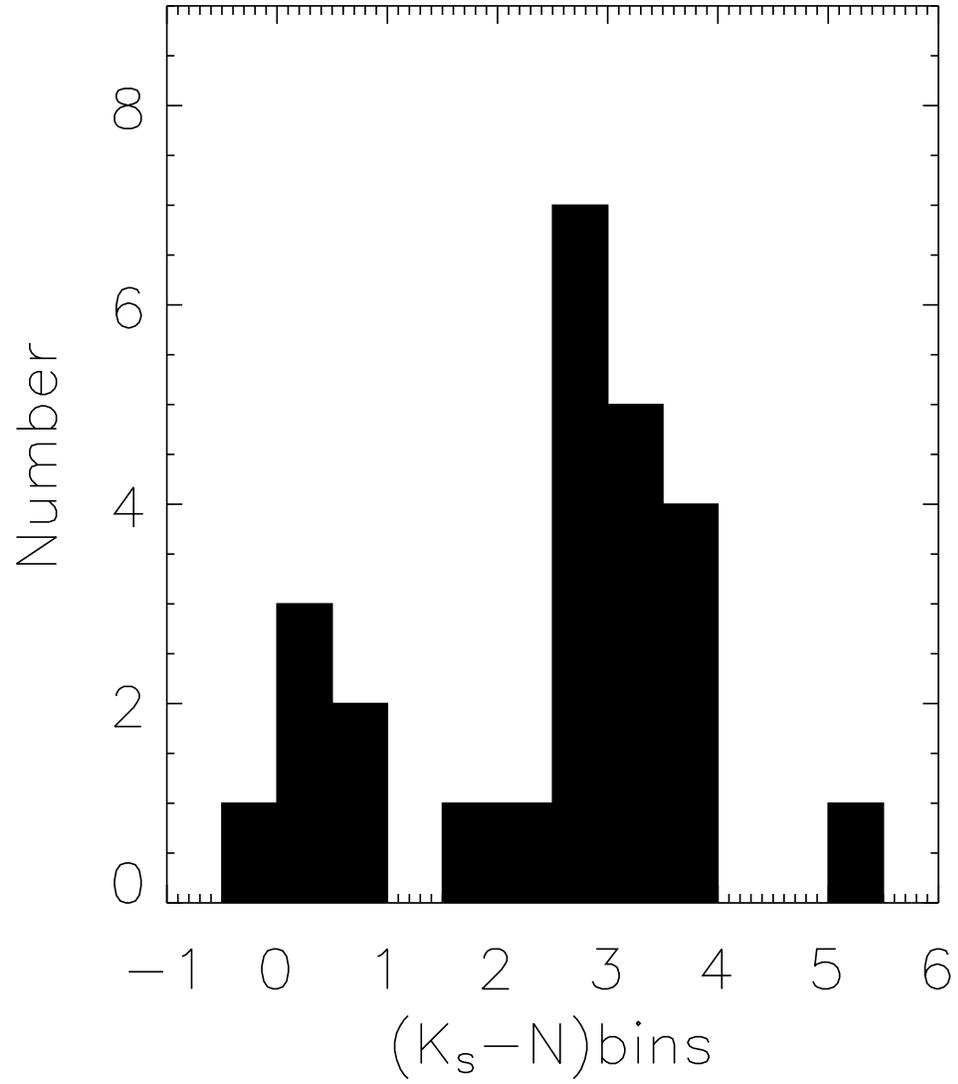}
\caption{Histogram of $(K_{s}-N)$ bins (binsize=0.5m) for objects in Figure 1a(excluding upper-limits). Notice the paucity of stars between $1.0m < (K_{s}-N) < 2.0m$.\label{fig2}}
\end{figure}
\clearpage

\begin{figure}
\plotone{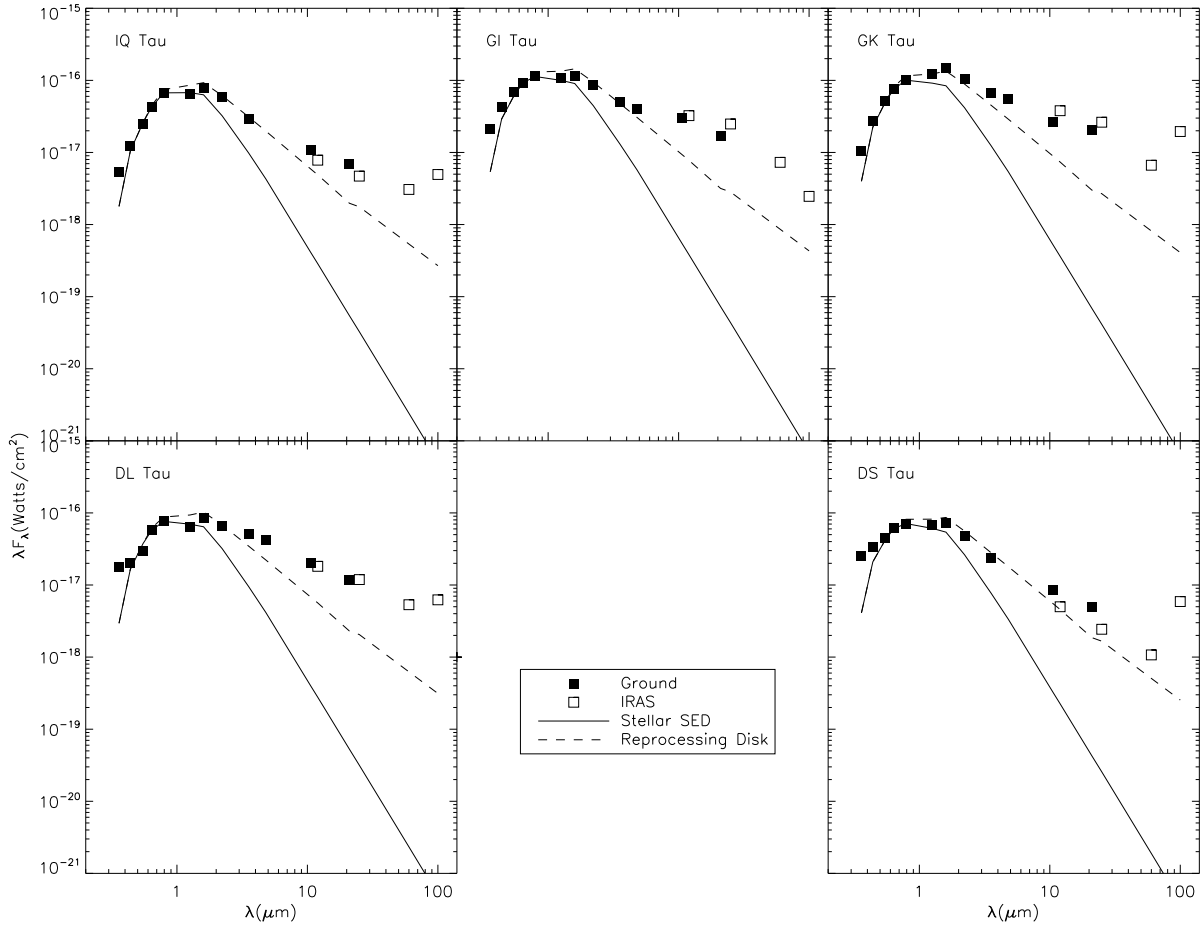}
\caption{These are SEDs of five objects with new M and Q-band observations. The observed fluxes are compared to a model stellar photosphere and an optically-thick, geometrically-thin disk model \citep{hillenbrand92}. \label{fig3}}
\end{figure}
\clearpage

\begin{figure}
\plotone{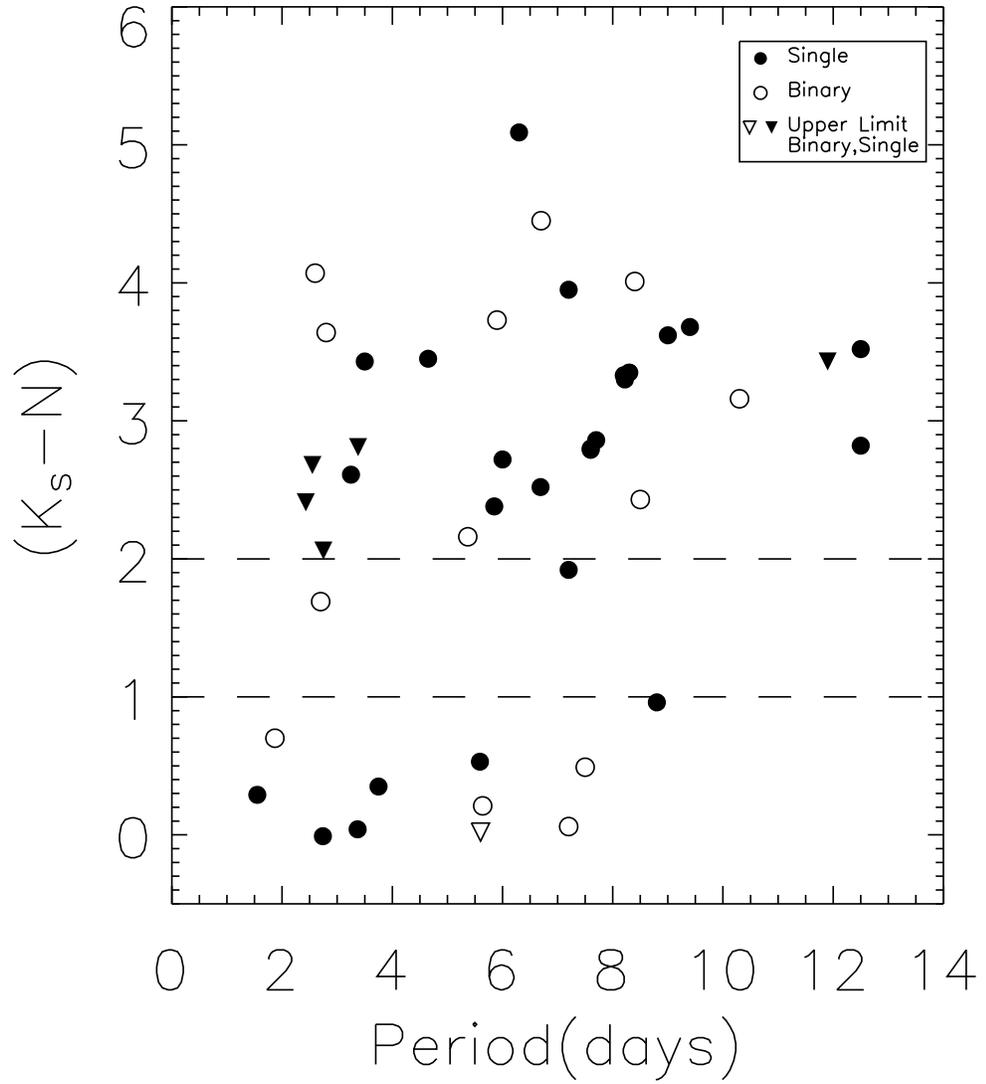}
\caption{This is a $(K_{s}-N)$ vs. period graph of single stars (from Figure 1a) and unresolved binaries with period data.\label{fig4}}
\end{figure}
\clearpage

\begin{deluxetable}{crccrc}
\tablecolumns{6}
\tablewidth{0pc}
\tablecaption{MAX Camera Observations of TAU-AUR objects at UKIRT\label{tbl-1}}
\tablehead{
\colhead{UT Date} & \colhead{HBC} & \colhead{Object Name} &
\colhead{Band} & \colhead{Exposure(sec)} & \colhead{Flux Standards} }
\startdata
1997 Feb 6 & 41      & \object{IQ Tau}    & N & 163.84 & HR 1370, $\mu$UMa, $\alpha$Lyr \\
-    & - & -   & Q & 98.24  & $\beta$And, HR 1370, $\mu$UMa, $\alpha$Lyr \\
-    & 56      & \object{GI Tau}    & M & 163.84 & $\beta$And, HR 1370, $\mu$UMa  \\
-    & - & -   & N & 81.92  & HR 1370,$\mu$UMa, $\alpha$Lyr  \\
-    & - & -   & Q & 98.24  & $\beta$And, HR 1370, $\mu$UMa, $\alpha$Lyr \\
-    & 57      & \object{GK Tau}    & M & 163.84 & $\beta$And, HR 1370, $\mu$UMa  \\
-    & - & -   & N & 81.92  & HR 1370,$\mu$UMa, $\alpha$ Lyr \\
-    & - & -   & Q & 98.24  & $\beta$And, HR 1370, $\mu$UMa, $\alpha$Lyr \\
-    & 58      & \object{DL Tau}    & N & 163.84 & HR 1370, $\mu$UMa, $\alpha$Lyr \\
-    & - & -   & Q & 98.24  & $\beta$And, HR 1370, $\mu$UMa, $\alpha$Lyr \\
-    & 75      & \object{DS Tau}    & N & 163.84 & HR 1370, $\mu$UMa, $\alpha$Lyr \\
-    & - & -   & Q & 98.24  & $\beta$And, HR 1370, $\mu$UMa, $\alpha$Lyr \\
-    & 384     & \object{FT Tau}    & N & 163.84 & HR 1370, $\mu$UMa, $\alpha$Lyr \\
-    & 397     & \object{TAP 41}    & N & 327.68 & HR 1370, $\mu$UMa, $\alpha$Lyr \\
-    & 405     & \object{V830 Tau}  & N & 327.68 & HR 1370, $\mu$UMa, $\alpha$Lyr \\
-    & 45      & \object{UX Tau A}  & N & 245.76 & HR 1370, $\mu$UMa, $\alpha$Lyr \\
1997 Feb 8 & 376     & \object{TAP 26}    & N & 614.40 & HR 1370, $\mu$UMa, $\alpha$Lyr \\
-    & 380     & \object{HD 283572} & N & 245.76 & HR 1370, $\mu$UMa, $\alpha$Lyr \\
-    & 382     & \object{LkCa 21}   & N & 614.40 & HR 1370, $\mu$UMa, $\alpha$Lyr \\
-    & 392     & \object{TAP 40}    & N & 614.40 & HR 1370, $\mu$UMa, $\alpha$Lyr \\
\enddata
\end{deluxetable}

\begin{deluxetable}{ccccccccccccc}
\tablewidth{0pt}
\tabletypesize{\scriptsize}
\tablecolumns{13}
\setlength{\tabcolsep}{0.05in}
\tablecaption{Data for single T Tauri stars\label{tbl-2}}
\tablehead{
\colhead{HBC} & 
\colhead{Object} & 
\colhead{Sp. Type} & 
\colhead{TTs} & 
\colhead{SED} & 
\colhead{P(days)} & 
\colhead{T$_{eff}(\degr$K)} & 
\colhead{$L_\star/L_\sun$} & 
\colhead{log(years)} & 
\colhead{$M_\star/M_\sun$} & 
\colhead{K$_{s}$} & 
\colhead{N} & 
\colhead{N-ref}}
\startdata
41 & \object{IQ Tau} & M0.5 & W & II & 12.5\tablenotemark{c} & 3785 & 0.65 & 5.81 & 0.38 & $7.78\pm0.02$ & $4.96\pm0.06$ & This work \\ 
56 & \object{GI Tau} & K6 & C & II & 7.20\tablenotemark{b} & 4205 & 0.85 & 6.00 & 0.58 & $7.89\pm0.02$ & $3.94\pm0.04$ & This work \\
57 & \object{GK Tau} & K7 & C & II & 4.65\tablenotemark{b} & 4060 & 1.17 & 5.67 & 0.45 & $7.47\pm0.02$ & $4.02\pm0.04$ & This work \\ 
58 & \object{DL Tau} & K7 & C & II & 9.40\tablenotemark{a} & 4060 & \nodata & \nodata & \nodata & $7.96\pm0.02$ & $4.28\pm0.03$ & This work \\
75 & \object{DS Tau} & K5 & C & II & 7.70\tablenotemark{c} & 4350 & 0.65 & 6.38 & 0.75 & $8.04\pm0.03$ & $5.18\pm0.09$ & This work \\ 
37 & \object{TAP 26} & K7 & W & III & 2.55\tablenotemark{b} & 4060 & 0.41 & 6.45 & 0.64 & $9.27\pm0.02$ & $>$\textit{6.59} & This work \\
380 & \object{HD 283572} & G5 & W & III & 1.55\tablenotemark{b} & 5770 & 6.50 & 6.70 & 1.85 & $6.87\pm0.02$ & $6.58\pm0.18$ & This work \\ 
382 & \object{LkCa 21} & M3 & W & III & 8.80\tablenotemark{a} & 3470 & 0.62 & 5.27 & 0.24 & $8.45\pm0.02$ & $7.49\pm0.25$ & This work \\
384 & \object{FT Tau} & C & \nodata & II & 8.30\tablenotemark{c} & \nodata & \nodata & \nodata & \nodata & $8.60\pm0.02$ & $5.25\pm0.07$ & This work \\ 
392 & \object{TAP 40} & K5 & W & III & 3.38\tablenotemark{b} & 4350 & 0.32 & 7.03 & 0.87 & $9.50\pm0.02$ & $>$\textit{6.69} & This work \\
397 & \object{TAP 41} & K7 & W & III & 2.43\tablenotemark{b} & 4060 & 0.50 & 6.28 & 0.59 & $8.85\pm0.02$ & $>$\textit{6.44} & This work \\ 
405 & \object{V830 Tau} & K7 & W & III & 2.75\tablenotemark{b} & 4060 & 0.68 & 6.02 & 0.52 & $8.42\pm0.02$ & $>$\textit{6.36} & This work \\
25 & \object{CW Tau} & K3 & C & II & 8.20\tablenotemark{b} & 4730 & 1.35 & 6.27 & 0.97 & $7.13\pm0.02$ & $3.80\pm0.15$ & C74 \\ 
32 & \object{BP Tau} & K7 & C & II & 7.60\tablenotemark{b} & 4060 & 0.95 & 5.79 & 0.47 & $7.74\pm0.02$ & $4.95\pm0.11$ & CS76 \\
33 & \object{DE Tau} & M2 & C & II & 7.60\tablenotemark{b} & 3580 & 0.81 & 5.23 & 0.26 & $7.80\pm0.02$ & $5.00\pm0.25$ & C74 \\ 
37 & \object{DG Tau} & K7-M0 & C & II & 6.30\tablenotemark{b} & 3950 & \nodata & \nodata & \nodata & $6.99\pm0.02$ & $1.90\pm0.05$ & C74 \\
38 & \object{DH Tau} & M1 & C & II & 7.20\tablenotemark{b} & 3720 & 0.68 & 5.75 & 0.34 & $8.18\pm0.03$ & $6.26\pm0.04$ & M97b \\ 
63 & \object{AA Tau} & K7 & C & II & 8.22\tablenotemark{b} & 4060 & 0.74 & 5.96 & 0.51 & $8.05\pm0.02$ & $4.75\pm0.25$ & C74 \\
65 & \object{DN Tau} & M0 & C & II & 6.00\tablenotemark{b} & 3850 & 0.91 & 5.66 & 0.37 & $8.02\pm0.02$ & $5.30\pm0.30$ & C74 \\ 
67 & \object{DO Tau} & M0 & C & II & 12.5\tablenotemark{c} & 3850 & 1.20 & 5.50 & 0.34 & $7.30\pm0.02$ & $3.78\pm0.05$ & C74 \\
74 & \object{DR Tau} & K7 & C & II & 9.00\tablenotemark{b} & 4060 & \nodata & \nodata & \nodata & $6.87\pm0.02$ & $3.25\pm0.15$ & C74 \\ 
77 & \object{GM Aur} & K3 & C & II & 11.9\tablenotemark{b} & 4730 & 0.83 & 6.63 & 1.03 & $8.28\pm0.02$ & $>$\textit{4.85} & C74 \\
79 & \object{SU Aur} & G2 & W & II & 3.50\tablenotemark{a} & 5860 & 10.7 & 6.42 & 2.31 & $5.99\pm0.02$ & $2.56\pm0.05$ & C80 \\ 
370 & \object{LkCa 4} & K7 & W & III & 3.37\tablenotemark{b} & 4060 & 0.85 & 5.86 & 0.49 & $8.32\pm0.02$ & $8.28\pm0.27$ & S90 \\
378 & \object{V819 Tau} & K7 & W & III & 5.59\tablenotemark{b} & 4060 & 0.81 & 5.89 & 0.50 & $8.42\pm0.02$ & $7.89\pm0.16$ & S90 \\ 
385 & \object{IP Tau} & M0 & W & II & 3.25\tablenotemark{b} & 3850 & 0.43 & 6.17 & 0.48 & $8.35\pm0.02$ & $5.74\pm0.06$ & S90 \\
388 & \object{TAP 35} & K1 & W & III & 2.74\tablenotemark{b} & 5080 & 1.40 & 6.73 & 1.37 & $8.30\pm0.02$ & $8.31\pm0.22$ & S90 \\ 
399 & \object{V827 Tau} & K7 & W & III & 3.75\tablenotemark{b} & 4060 & 0.89 & 5.83 & 0.48 & $8.23\pm0.02$ & $7.88\pm0.16$ & S90 \\
419 & \object{LkCa 15} & K5 & W & II & 5.85\tablenotemark{b} & 4350 & 0.74 & 6.27 & 0.72 & $8.16\pm0.02$ & $5.78\pm0.11$ & S90 \\ 
429 & \object{V836 Tau} & K7 & W & III & 6.69\tablenotemark{b} & 4060 & 0.47 & 6.33 & 0.60 & $8.60\pm0.02$ & $6.08\pm0.09$ & S90 \\
\enddata
\tablenotetext{a}{\citet{bouvier93}}
\tablenotetext{b}{\citet{bouvier95}}
\tablenotetext{c}{\citet{osterloh96}}
\tablecomments{Spectral Type, SED class, $T_{eff}$ and $L_{\star}$ are from \citet{KH95}. TTS type was taken from \citet{strom89}. log(years) and $M_{\star}$ derived from tracks of \citet{DM97}. N-Ref. C74 - \citet{C74}, C80 - \citet{C80}, M97b - \citet{meyer97b} and S90 - \citet{S90}. 3$\sigma$ upper-limits are quoted for non-detections.}
\end{deluxetable}

\clearpage

\begin{deluxetable}{ccccccccccccccccc}
\tablecolumns{17}
\tablewidth{0pt}
\rotate
\tabletypesize{\scriptsize}
\tablecaption{Photometric Data for SEDs.\label{tbl-3}}
\tablehead{
\colhead{Object} & 
\colhead{U} & 
\colhead{B} &
\colhead{V} & 
\colhead{R} & 
\colhead{I} & 
\colhead{J} & 
\colhead{H} & 
\colhead{K$_s$} & 
\colhead{L} & 
\colhead{M} &
\colhead{N} & 
\colhead{12$\micron$} & 
\colhead{Q} &
\colhead{25$\micron$} & 
\colhead{60$\micron$} & 
\colhead{100$\micron$} }
\startdata
GI Tau	& 15.28	& 14.79	& 13.21	& 12.15	& 11.06	& 9.42	& 8.46	& 7.89	& 6.83	& $6.11\pm0.21$ & $3.94\pm0.04$ & 3.43 & $2.28\pm0.09$ & 1.30 & -0.22 & -0.70 \\
GK Tau	& 14.56	& 14.01	& 12.54	& 11.58	& 10.62	& 9.02	& 8.02	& 7.47	& 6.46	& $5.72\pm0.15$ & $4.02\pm0.04$ & 3.22 & $2.07\pm0.10$ & 1.23 & -0.12 & -2.95 \\
DL Tau	& 13.91	& 14.26	& 13.12	& 11.85	& 10.89	& 9.73	& 8.63	& 7.96	& 6.76	& 6.04 & $4.28\pm0.03$ & 4.02 & $2.66\pm0.12$ & 2.09 & 0.12  & -1.71 \\
IQ Tau	& 15.43	& 14.99	& 13.45	& 12.28	& 11.11	& 9.74	& 8.74	& 7.78	& 7.39	& \nodata  & $4.96\pm0.06$ & 4.94 & $3.25\pm0.21$ & 3.10 & 0.72  & -1.46 \\
DS Tau	& 13.11	& 13.35	& 12.37	& 11.56	& 10.80	& 9.59	& 8.75	& 8.04	& 7.57	& \nodata  & $5.18\pm0.09$ & 5.41 & $3.61\pm0.30$ & 3.81 & 1.86  & -1.65 \\
\enddata
\tablecomments{Values from this study have uncertainties quoted. UBVRI and JHLM are from KH95. K$_{s}$ is from the 2MASS point source catalog. 12$\micron$, 25$\micron$, 60$\micron$ and 100$\micron$ values are from the IRAS point source catalog \citep{beichman88}.}
\end{deluxetable}

\begin{deluxetable}{ccccccccc}
\tablecolumns{9}
\tablewidth{0pc}
\tablecaption{Data for Binaries with Periods and N-band magnitudes.\label{tbl-4}}
\tablehead{
\colhead{HBC} & \colhead{Object} & \colhead{P(days)} &
\colhead{Sep($\arcsec$)} & \colhead{$T_{eff}(\degr$K)} & \colhead{$L_{\star}/L_{\sun}$} &
\colhead{K$_{s}$} & \colhead{N} & \colhead{N-ref.} }
\startdata
29  & \object{V410 Tau} & 1.87 & 0.12\tablenotemark{a} & 4730 & 2.14 & 7.63 & 6.93 & S90	\\
35  & \object{T Tau} & 2.80 & 0.71\tablenotemark{a} & 5250 & 8.91 & 5.33 & 1.69 & G91 \\
36  & \object{DF Tau} & 8.50 & 0.09\tablenotemark{b} & 3470 & 1.60 & 6.73 & 4.30 & C74	\\
39  & \object{DI Tau} & 7.50 & 0.12\tablenotemark{b} & 3850 & 0.62 & 8.39 & 7.90 & M97b \\ 
43  & \object{UX Tau A} & 2.70 & 2.70\tablenotemark{c} & 4900 & 1.35 & 7.55 & $5.86\pm0.10$& This work\\
45  & \object{DK Tau} & 8.40 & 2.53\tablenotemark{b} & 4060 & 1.32 & 7.10 & 3.09 & CS76 \\
50  & \object{XZ Tau} & 2.60 & 0.31\tablenotemark{a} & 3470 & 0.71 & 7.29 & 3.22 & CS76 \\
54  & \object{GG Tau} & 10.3 & 0.29\tablenotemark{a} & 4060 & 1.50 & 7.36 & 4.20 & C74	\\
66  & \object{HP Tau} & 5.90 & 0.02\tablenotemark{b} & 4730 & 1.30 & 7.63 & 3.90 & C74	\\
68  & \object{VY Tau} & 5.37 & 0.66\tablenotemark{b} & 3850 & 0.47 & 8.96 & 6.80 & SP95 \\
76  & \object{UY Aur} & 6.70\tablenotemark{*} & 0.88\tablenotemark{a} & 4060 & 2.00 & 7.24 & 2.79 & KHL97\\
368 & \object{LkCa 3} & 7.20 & 0.49\tablenotemark{a} & 3720 & 1.66 & 7.42 & 7.36 & S90 \\
379 & \object{LkCa 7} & 5.64 & 1.05\tablenotemark{c} & 4060 & 0.89 & 8.26 & 8.05 & S90	\\
420 & \object{IW Tau} & 5.60 & 0.27\tablenotemark{b} & 4060 & 0.87 & 8.28 & $>$\textit{8.26} & SMVMH01 \\
\enddata
\tablenotetext{*}{\citet{osterloh96}. All other periods are from \citet{bouvier95}}
\tablenotetext{a}{\citet{ghez93}}
\tablenotetext{b}{\citet{simon95}}
\tablenotetext{c}{\citet{leinert93}}
\tablecomments{UX Tau's companions were unresolved in the N-band. We are unsafely assuming that UX Tau A is the dominant source of N-band flux(See section 4.3).\\ $T_{eff}$ and $L_\star$ values are from \citet{KH95}. K$_{s}$ magnitudes are from the 2MASS point source catalog. C74 - \citet{C74}, CS76 - \citet{CS76}, S90 - \citet{S90}, G91 - \citet{ghez91}, SP95 - \citet{SP95}, M97b - \citet{meyer97b}, KHL97 - \citet{KHL97}, SMVMH01 - \citet{stassun01}. 3$\sigma$ magnitude upper-limits are quoted for non-detections.}
\end{deluxetable}

\end{document}